# Defect–Charge–Driven 90° Switching in HfO$_2$

Muting Xie[a,b], Hongyu Yu[a,b], Zhihao Dai[a,b], Yingfen Wei[c], Changsong Xu[a,b*], and Hongjun Xiang[a†]

[a]*Key Laboratory of Computational Physical Sciences (Ministry of Education), Institute of Computational Physical Sciences, State Key Laboratory of Surface Physics, and Department of Physics, Fudan University, Shanghai 200433, China*

[b]*Hefei National Laboratory, Hefei 230088, China.*

[c]*State Key Laboratory of Integrated Chips and Systems, Frontier Institute of Chip and System and, Zhangjiang Fudan International Innovation Center, Fudan University, Shanghai, China*

*E-mail: csxu@fudan.edu.cn

†E-mail: hxiang@fudan.edu.cn

**Abstract**

Hafnium dioxide (HfO$_2$) is a CMOS-compatible ferroelectric showing both 180° and 90° switching, yet the microscopic nature of the 90° pathway remains unresolved. We show that the 90° rotation pathway, negligible in pristine HfO$_2$, becomes dominant under $\vec{E}$ // [111] when induced by charged oxygen vacancies. This pathway is more fatigue-resistant than the 180° reversal pathway, while delivering the same polarization change along [111] ($2Pr = 60\ \mu C/cm^2$). This charge–driven switching arises from two factors: the crystal geometry of HfO$_2$ and the intrinsic nature of rotational pathways, the latter suggesting a possible general tendency for defect charge to bias rotation over reversal in ferroelectrics. Together these findings reveal a pathway-level origin of fatigue resistance and establish defect charge as a general control parameter for polarization dynamics.

**Introduction**

Hafnium dioxide (HfO$_2$)–based ferroelectrics are promising candidates for advanced complementary metal–oxide–semiconductor (CMOS) technology, yet their switching physics remains complex [1–4]. Experimentally, multiple crystallographic orientations are accessible, and both 180° and 90° switching events have been observed [5–8]. According to modern polarization theory, the pathways connecting ferroelectric states are indispensable to a complete description of ferroelectric behavior [9,10]. Nevertheless, the fundamental distinction between 180° and 90° processes—their occurrence conditions, microscopic differences, and relative stability—remains unresolved, with the microscopic nature of 90° rotation being particularly unclear.

In HfO$_2$, substantial efforts have been devoted to understanding polarization switching [11–15]. 180° reversal has been analyzed in detail, both at the domain level and in terms of microscopic pathways [12,16–21]. By contrast, studies of 90° rotation have focused mainly on domain motion [13,16,22–24], leaving the microscopic pathway largely unexplored. Experiments frequently observe both [001] and [111] orientations, which sustain nearly equal polarization magnitudes [25–27], yet display divergent switching behavior. These observations raise two questions: whether the differences arise from distinct processes, and what determines whether a given orientation switches by 180° or by 90°.

Oxygen vacancies are ubiquitous in oxide ferroelectrics and critically influence switching kinetics through mechanisms such as nucleation–growth [28–30]. As intrinsic defects, they are also implicated in fatigue, for instance through domain-wall pinning [31–33]. In HfO$_2$, charged vacancies are known to stabilize the orthorhombic ferroelectric phase [34–36], but their impact on switching pathways—equally essential to ferroelectric behavior—remains unresolved. Here we show that vacancy charge, coupled with crystallographic orientation, decisively selects the switching mode.

In this Letter, we show that vacancy charge, in concert with crystallographic orientation, resolves the long-standing uncertainty between 180° and 90° switching in HfO$_2$. Charged oxygen vacancies turn the otherwise negligible 90° rotation into the dominant pathway under $\vec{E}$ // [111], biasing pathway competition through two factors: the crystal geometry of HfO$_2$ and the intrinsic nature of rotational switching. We further find that 90° rotation persists more robustly under cyclic fields than 180° reversal, revealing a pathway-level origin of fatigue resistance. Together these results establish a microscopic picture in which defect charge governs ferroelectric pathway selection, providing a general principle for controlling polarization dynamics.

**Method**

For our computational approach, DFT was used for static properties, including energies, polarization, and charge distribution. Machine-learning models, in contrast, were employed for dynamic processes such as field-driven molecular dynamics and switching pathways. We trained separate models for charge-neutral and charged systems with DREAM [37], a framework that predicts interatomic potentials and Born effective charges (BECs). The datasets comprised pristine, single-vacancy, and double-vacancy configurations. Vacancies were sampled at both polar and nonpolar oxygen sites, with polar sites used as the default reference, as they are energetically favored. The charge-neutral model was trained on 13,487 structures (force error 0.06 eV/Å per atom, BEC error of 0.10 e. The charged-vacancy model, trained on a larger set of 15,885 structures, attained slightly better force accuracy (0.04 eV/Å, 0.07 e). Further details of the DFT setup, dataset composition, model benchmarks, and vacancy energetics are provided in the Supplementary Materials [38].

**Results**

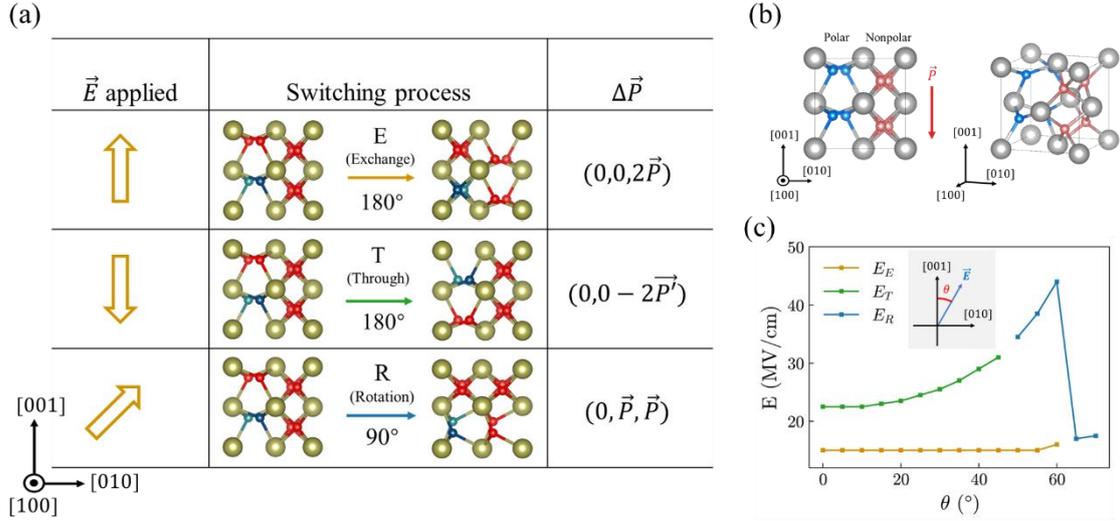

**Fig.1** Switching pathways in pristine system. **(a)** Polarization-switching pathways versus field direction, with corresponding changes $\Delta \vec{P}$. Different colors indicate atomic correspondence. **(b)** Unit cell of the polarized $Pca2_1$. Oxygen atoms grouped into polar (blue) and nonpolar (red) types. **(c)** Coercive field $\vec{E}$ as a function of field angle θ for E-, T-, and R-paths in a Pristine cell.

The 90° rotation pathway (R-path) in pristine HfO₂ emerges as a candidate pathway, yet its stringent angular constraint renders it negligible for sustained ferroelectric switching. Among the three polarization-switching pathways—two 180° reversals (E- and T-paths) and one 90° rotation (R-path)—the field direction dictates which mode is activated. When the field is parallel to [001] (antiparallel to the polarization), the E-path is activated through exchange between the polar and nonpolar oxygen sublattices [18]. When the field is antiparallel to [001], the T-path occurs via polar oxygen atoms passing through the Hf plane [12]. Tilting the field toward [010] rotates the polarization, giving the R-path. Figure 1(c) shows that while the coercive field of the R-path ($E_R \approx 17$ MV/cm) is close to that of the E-path ($E_E \approx 16$ MV/cm), it is accessible only for tilt angles $\theta > 60°$ in the [001]–[010] plane. Such a restricted window makes the R-path effectively irreversible, leaving the E-path as the operative mode in both [001] and [111] orientations.

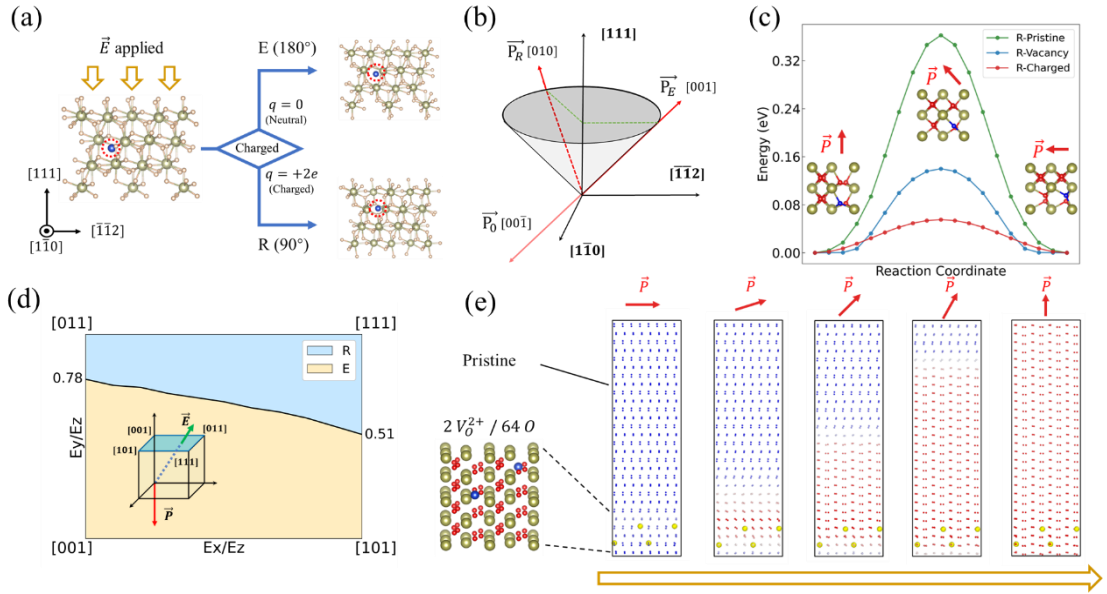

**Fig.2** Switching pathways in systems with oxygen vacancies. **(a)** For $\vec{E}$ // [111], the switching pathway depends on vacancy charge state ($V_O$ or $V_O^{2+}$). Blue sphere marks the vacancy. **(b)** Polarization vectors for 180° reversal and 90° rotation, both giving the same $\Delta P$ along [111]. **(c)** R-path energy barriers for Pristine (no vacancy), Vacancy (one $V_O$ per unit cell), and Charged (one $V_O^{2+}$ per unit cell). The intermediate state resembles the Pmn2₁ phase, a ferroelectric phase that preserves polarization during the transition. **(d)** Phase diagram of ferroelectric switching for two $V_O^{2+}$ among 64 O atoms. **(e)** Nucleation–growth process driven by $V_O^{2+}$. blue/red: opposite polarization states; yellow: vacancy positions fixed by crystal periodicity. Vacancies are confined to the lower 10 Å of the 70 Å supercell.

Although negligible in pristine systems, charged oxygen vacancies render the 90° R-path feasible for $\vec{E}$ // [111] [Fig. 2(a)]. The energy barrier for the R-path in a pristine 4 Hfs and 8 Os cell is 0.36 eV. Introducing a single neutral vacancy $V_O$ lowers it to 0.14 eV, and a single charged vacancy $V_O^{2+}$ further reduces it to 0.06 eV [Fig. 2(c)]. Under the same geometry as Fig. 2(a), this sharp reduction in barrier leads to distinct switching modes: $V_O$ drives a 180° E-path, whereas $V_O^{2+}$ activates the 90° R-path. Crucially, the 90° rotation does not reduce the usable polarization: along [111] it yields the same change as the 180° reversal, $2Pr = 60 \ \mu C/cm^{\wedge}2$ [Fig. 2(b)]. This demonstrates that the vacancy charge state plays a decisive role in selecting the switching pathway.

Two factors make the R-path with $V_O^{2+}$ not only accessible but also dominant under $\vec{E}$ // [111]: its broad angular tolerance and a vacancy-driven nucleation–growth mechanism. As shown in Fig. 2(d), successive R-path switching is sustained within an acceptance cone of about 10°–30°, depending on the deviation direction of the applied field. In the range shown in Fig. 2(d), the [001] component of the field remains the largest. This dominance is the necessary condition for the R-path to be accessible, because the R-path must allow activation from both [001] and [010] orientations. The second factor, nucleation–growth, relaxes the spatial and density requirements for $V_O^{2+}$ s. As seen in Fig. 2(e), a rotated domain nucleates in the lower 10 Å of a 70 Å supercell, initiated by localized vacancies, and then propagates through the crystal along its polarization front. Thus, even a localized density as low as two vacancies among 64 O atoms is already sufficient for the R-path to propagate through the crystal and outcompete the E-path.

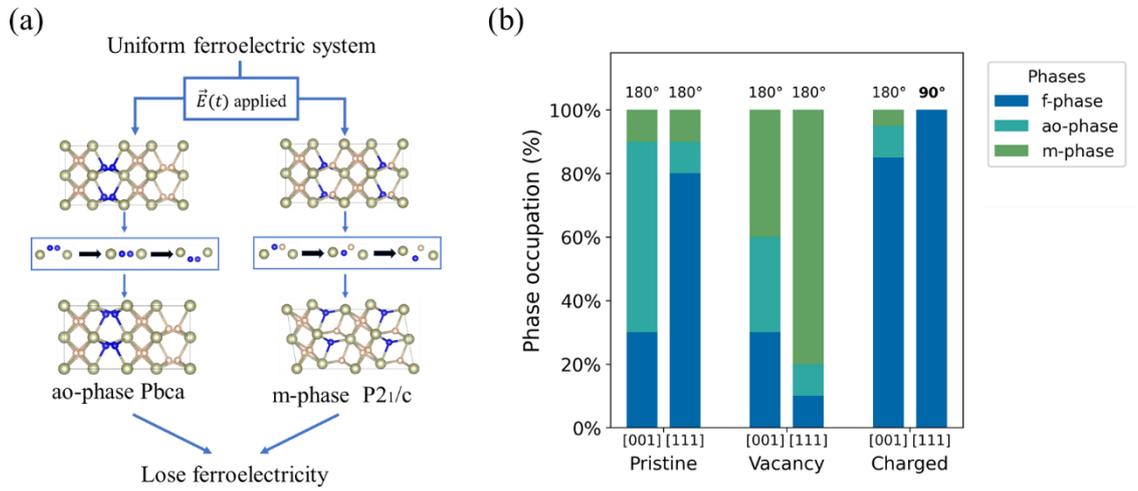

**Fig. 3** Fatigue in ferroelectric HfO₂ from molecular dynamics at 300 K. Simulations used a 100Å×20Å×20Å supercell under alternating electric fields along [001] or [111], with three states: Pristine, Vacancy ($V_O$ concentration 0.6%), and Charged ($V_O^{2+}$ concentration 0.6%). **(a)** Fatigue process observed in simulation. **(b)** Phase diagrams after 20 cycles of alternating fields.

R-paths exhibit enhanced fatigue resistance, a key advantage for long-term operation. To evaluate this, we tracked structural transformations under cyclic fields. Fatigue originates from two T-path–related processes [Fig. 3(a)]. First, inter-cell nonuniform T-paths yield the ao-phase (Pbca) [39,40] , effectively acting as pinning centers. Second, intra-cell partial pass-through of polar oxygens (~50%) produces the m-phase ($P2_1/c$). Both processes share the critical "pass-through" step, which can be induced by E-paths [20]. Because R-paths do not involve this step, replacing E-paths with R-paths suppresses these fatigue-inducing transformations. Fig. 3(b) compares fatigue resistance across field orientations and defect states. For $\vec{E}$ // [111], the Charged system follows the R-path and remains fatigue-free over 20 cycles, whereas all others follow E-paths and exhibit fatigue to varying degrees within 10 cycles. Three additional trends emerge: fields along [111] suppress ao-phase formation; neutral vacancies promote fatigue by accelerating unfavorable phase transitions relative to Pristine; and even under [001] fields, charged vacancies can still mitigate fatigue. These trends are consistent with experimental reports [39,40]; their broader implications are discussed below.

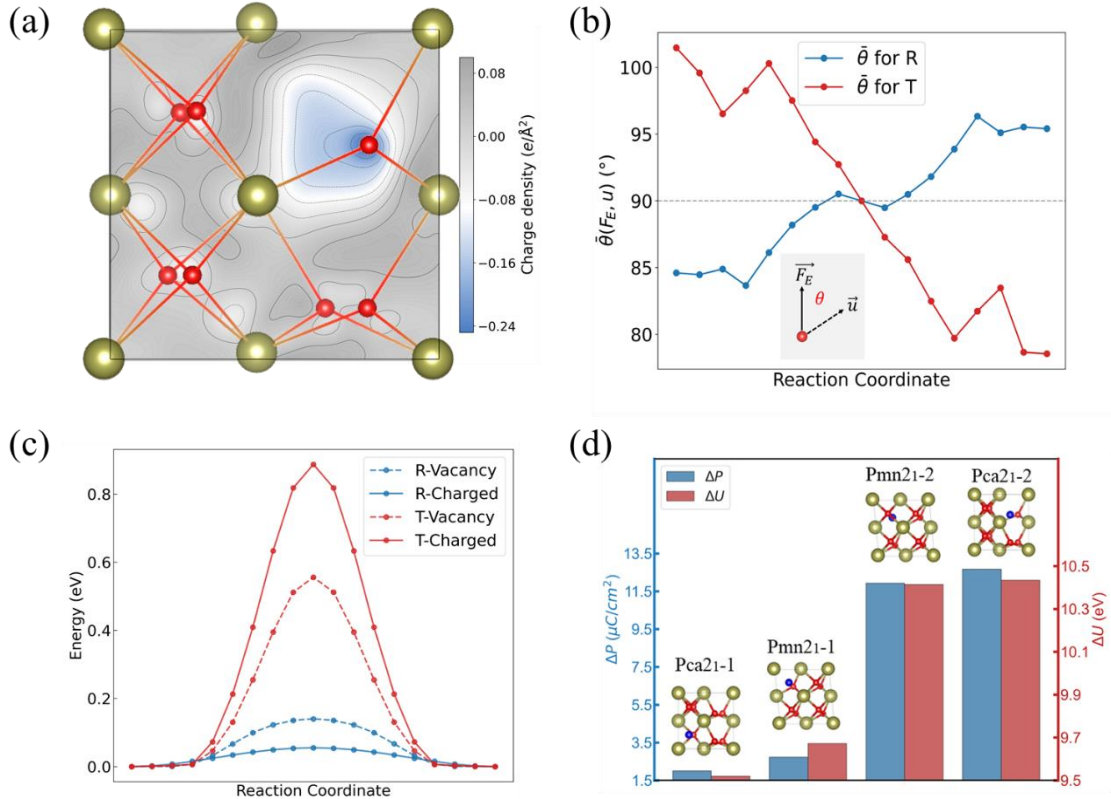

**Fig. 4** Influence of defect charge q. (a) Defect charge q, the charge-density difference between Charged and Vacancy systems, integrated along z. (b) Weighted mean included angle $\theta(\vec{F_q}, \vec{u})$ between the electrostatic force $\vec{F_q}$ induced by q and the oxygen displacement $\vec{u}$ during switching. (c) Switching barriers for R-path and T-path in Vacancy and Charged systems. (d) Reduction of polarization (ΔP) and energy (ΔU) upon charging oxygen vacancies. illustrated for different vacancy configurations in $Pca2_1$ and $Pmn2_1$ states.

We attribute the R-path preference induced by charged vacancies to two factors— one specific to the crystal geometry of $HfO_2$ and one rooted in the generic nature of "rotation"— both ultimately traced to the defect charge q, defined as the net charge difference between Charged and Vacancy systems [Fig. 4(a)]. The crystal geometry dictates that the oxygen displacement during switching can align with or oppose q-induced electrostatic force. For contrast, we use the T-path—the origin of fatigue [Fig. 3(a)]—as a reference. This correlation is quantified via the included angle $\theta(\vec{F_E}, \vec{u})$, where $\vec{F_E}$ is the electrostatic force and $\vec{u}$ is the oxygen displacement. For the R-path, $\theta(\vec{F_E},\vec{u}) < 90°$ before the intermediate state, so the internal field assists the atomic motion; the T-path exhibits the opposite trend [Fig. 4(b)]. That is, geometry directs the defect-charge–induced field to drive the R-path but block the T-path.

The second factor arises from the "90° rotation" character of the R-path, a feature potentially generic to other ferroelectrics. A key role of defect charge q is its ability to stabilize polarized states. Its distribution responds to the existing polarization, tending to compensate it and thus may reduce the electrostatic energy. As illustrated in Fig. 4(d), charging oxygen vacancies reduces both the polarization (ΔP) and the energy (ΔU) between distinct vacancy configurations, with ΔU and ΔP showing a clear positive correlation. For switching pathways with a paraelectric intermediate, such stabilization deepens the initial state and increases the barrier; in $HfO_2$, this raises the T-path barrier by 0.33 eV [Fig. 4(c)]. In contrast, a rotation pathway retains polarization throughout the transition, making it inherently more favorable under defect charge; here, the R-path proceeds through a ferroelectric intermediate ($Pmn2_1$), lowering its barrier by 0.08 eV instead [Fig. 4(c)]. Altogether, these results indicate that defect charge stabilizes rotation, as polarization is naturally preserved through the intermediate state.

**Discussion**

These results show how defect charge biases the system toward 90° rotation in $HfO_2$. Normally suppressed by angular constraints, it becomes the dominant mode under [111] fields when vacancies carry charge. The reason is twofold: charge aligns with the crystal geometry to guide oxygen motion, and it lowers the barrier by stabilizing a ferroelectric intermediate. By displacing the E-path, the R-path avoids the fatigue-inducing transformations associated with the T-path, explaining its superior endurance.

Experiments often report both 180° and non-180° switching events [5,6]. Our framework explains this coexistence as orientation and defect charge biasing the system between E- and R-paths. This picture also accounts for the appearance of fatigue-related

ao- [39,40] and m-phases [41] , which stem from T-path pass-through. It further clarifies the superior ferroelectricity of [111]-oriented films [26] , where rotation becomes accessible under defect charge. Direct evidence for charged-defect–induced rotation remains elusive, but the suppression of the m-phase under $CeO_{2-x}$ capping [41] is consistent with our mechanism.

Defect-charge control of rotation carries broader implications. In $HfO_2$, the ability of charged vacancies to stabilize rotation suggests a practical strategy: exploiting, rather than eliminating, intrinsic defects to improve device reliability. More broadly, because rotation-type pathways preserve polarization through a ferroelectric intermediate, defect charge could likewise favor rotation in other ferroelectrics. This raises the possibility that defect–charge–mediated pathway selection may represent a general mechanism for tuning polarization dynamics.

## Summary

In summary, charged oxygen vacancies govern the switching pathway in $HfO_2$. While 90° rotation is negligible in pristine systems, it becomes dominant—and the more fatigue-resistant pathway—once stabilized under [111] orientation by defect charge. Such behavior arises from the crystal geometry of $HfO_2$ and the intrinsic nature of 90° rotation. In $HfO_2$, this mechanism offers a practical strategy of exploiting, rather than eliminating, defects to improve device reliability. More broadly, defect-charge–mediated pathway selection may constitute a general principle of polarization dynamics in ferroelectrics.

## Acknowledgements

We acknowledge financial support from the National Key R&D Program of China (No. 2022YFA1402901), NSFC (grants No. 12188101, 12274082), Shanghai Science and Technology Program (No. 23JC1400900), the Guangdong Major Project of the Basic and Applied Basic Research (Future functional materials under extreme conditions--2021B0301030005), Shanghai Pilot Program for Basic Research—FuDan University 21TQ1400100 (23TQ017), the robotic AI-Scientist platform of Chinese Academy of Science, and New Cornerstone Science Foundation. C.X.

also acknowledges support from the Shanghai Science and Technology Committee (Grant No. 23ZR1406600), Shanghai Education Committee (Grant No. 24KXZNA01), Innovation Program for Quantum Science and Technology (2024ZD0300102), and the Xiaomi Young Talents Program.